\pdfoutput=1

\documentclass[lettersize,journal]{IEEEtran}
\usepackage{amsmath,amsfonts}
\usepackage{algorithmic}
\usepackage{algorithm}
\usepackage{array}
\usepackage[caption=false,font=normalsize,labelfont=sf,textfont=sf]{subfig}
\usepackage{textcomp}
\usepackage{stfloats}
\usepackage{url}
\usepackage{verbatim}
\usepackage{graphicx}
\usepackage{cite}
\usepackage{booktabs}
\begin{document}

\title{MVC: A Multi-Task Vision Transformer Network for COVID-19 Diagnosis from Chest X-ray Images}

\author{Huyen Tran, 
        Duc Thanh Nguyen,~\IEEEmembership{Member,~IEEE,} 
        John Yearwood
\thanks{Huyen Tran, Duc Thanh Nguyen, and John Yearwood are with the School of Information Technology, Deakin University, 75 Pigdons Road, Waurn Ponds, VIC 3216, Australia (e-mail: \{trhu,duc.nguyen,john.yearwood\}@deakin.edu.au).}}



\maketitle

\begin{abstract}

Medical image analysis using computer-based algorithms has attracted considerable attention from the research community and achieved tremendous progress in the last decade. With recent advances in computing resources and availability of large-scale medical image datasets, many deep learning models have been developed for disease diagnosis from medical images. However, existing techniques focus on sub-tasks, e.g., disease classification and identification, individually, while there is a lack of a unified framework enabling multi-task diagnosis. Inspired by the capability of Vision Transformers in both local and global representation learning, we propose in this paper a new method, namely Multi-task Vision Transformer (MVC) for simultaneously classifying chest X-ray images and identifying affected regions from the input data. Our method is built upon the Vision Transformer but extends its learning capability in a multi-task setting. We evaluated our proposed method and compared it with existing baselines on a benchmark dataset of COVID-19 chest X-ray images. Experimental results verified the superiority of the proposed method over the baselines on both the image classification and affected region identification tasks.

\end{abstract}


\section{Introduction}
\label{sec:introduction}

\IEEEPARstart{C}{hest X-ray} (CXR) imaging is a common method for diagnosing lung related diseases. Performing CXR is relatively cheap and quick, yet effective as a CXR image captures sufficient details about the condition of a patient's lungs. Well-trained doctors and radiologists can study known patterns of lung diseases from text books and then practice diagnosis on CXR databases. However, identifying disease patterns from CXR images is challenging, especially when nodules in the patterns are small and/or appear at positions that align with organs.

CXR image examination is critical to deciding correct treatments and has often been performed manually. However, during the outbreak of the COVID-19 pandemic, where the sheer number of patients outnumbers the number of doctors and radiologists, manual diagnoses are too expensive and time consuming for even a fraction of patients. 

Literature has shown that it is possible to automate parts of the diagnosis process using computer-based medical image analysis techniques~\cite{litjens2017survey,ccalli2021deep}. Specifically, computer-based algorithms can be used to predict the health condition of patients based on their CXR images. Those algorithms are trained to extract complex patterns from CXR data using training signals provided by radiologists and experts. After being trained and validated carefully, the algorithms are applied to diagnose lung related diseases from CXR images. There are three main challenges to be addressed: (i) CXR patterns are different from normal images and unclear even to human experts. (ii) Nodules in a CXR image vary in their shapes and sizes and can be found at arbitrary locations. (iii) It is important to point out regions affected by the disease. These regions are evidence to help doctors and radiologists consolidate their diagnoses.

With recent developments of hardware devices that enable complex computations and availability of large-scale datasets that provide abundant training data, deep learning algorithms have been widely applied to medical image analysis and achieved impressive outcomes in many tasks including image classification~\cite{narin2021automatic}, segmentation~\cite{Kamal_JBHI_2022}, and registration~\cite{Lee_JMBE_2021}. CXR research has also benefited from such high capacity deep neural network architectures to learn and extract complex patterns from CXR data. 

A common practice of deep learning-based CXR image analysis is to adopt a pre-trained convolutional neural network (CNN), trained on large-scale imagery datasets such as ImageNet~\cite{imagenet_cvpr09}, then fine-tune it on a CXR dataset for a specific diagnosis task. However, this approach shows several limitations. Firstly, pre-trained CNN architectures are not specifically designed to work with CXR patterns covering bilateral involvement, peripheral and lower zone dominance of ground glass opacities, and patchy consolidations~\cite{cozzi2020chest}. Secondly, CNNs often make use of simple feature concatenation and pooling operations to fuse local features into global ones. However, feature interactions between different spatial regions are not taken into account. In addition, existing methods focus on individual diagnosis tasks, i.e., a CNN is designed per diagnosis task, while lacking a unified framework to support multiple tasks, e.g., disease recognition and affected region identification. Different tasks may have dual and correlation relationships, leveraging the overall performance. 

In this paper, we propose a unified framework to address the above issues. Our method is designed to learn both local patterns and global correlations between them from CXR images while simultaneously performing both chest X-ray image classification and affected region identification. Our method is built upon the Vision Transformer (ViT)~\cite{dosovitskiy2020image} but extends the learning capability via incorporation of both local and global information, and multi-task learning. To this end, we make the following contributions.
\begin{itemize}
    \item Multi-task Vision Transformer (MVC), a network architecture that simultaneously enables two common diagnosis tasks from CXR images: disease recognition and affected region identification. Our architecture can help radiologists determine a disease from a CXR image and, at the same time, localise image regions relevant to the disease. This region-based information is important as it would help consolidate and support decision making to medical treatments. To the best of our knowledge, MVC is the first method tackling these tasks explicitly and simultaneously.
    \item Effective incorporation of local patterns and global correlations via self-attention and multi-task learning. It is proven that such an incorporation improves the learning capability of the architecture, while multi-task learning further boosts up the performance of each individual task. 
    \item Extensive experiments on a benchmark dataset to validate and compare our method with existing baselines.
\end{itemize}

The remainder of our paper is organised as follows. Section~\ref{sec:related_work} discuss related work. Section~\ref{sec:proposed_method} presents our proposed method, including the MVC's architecture and its training. We report experimental results in Section~\ref{sec:exp}. Section~\ref{sec:discussion} summarises our paper with remarks and discusses future work.

\section{Related Work}
\label{sec:related_work}

\subsection{CNN-based methods}

Literature in deep learning-based CXR image analysis has mainly focused on applying CNN architectures to learn complex patterns from CXR data for different diagnosis tasks~\cite{litjens2017survey,ccalli2021deep}. Since training of CNNs requires a lot of training data, one often adopts architectures, pre-trained on public and large-scale datasets, such as ImageNet~\cite{imagenet_cvpr09}, and customise them, for instance by replacing the last layer, to fit with a medical task, such as a disease recognition task. These CNNs model, after being customised, are fine-tuned on a smaller domain-specific dataset relevant to the task. Although the CXR image domain is very different from those in natural images in ImageNet dataset, it is empirically shown that shallow layers in pre-trained models could still capture useful information for downstream applications. 

Example architectures commonly adopted in CXR image analysis includes VGG~\cite{simonyan2014very} in~\cite{apostolopoulos2020COVID,shelke2021chest}, ResNet~\cite{he2016deep} in~\cite{narin2021automatic,sitaula2021attention,sethy2020detection},
DenseNet~\cite{huang2017densely} in~\cite{singh2021densely,rajpurkar2017chexnet}, and combination of ResNet and DenseNet in~\cite{Chen_BSPC_2019}. To summarise this trend, the work in~\cite{nayak2021application} compared several popular pre-trained models for automated COVID-19 screening from CXR images. However, domain shift is a well-known issue of this approach, and becomes more severe when there are limited labelled data in the CXR domain. For instance, as shown in~\cite{roberts2021common,zech2018variable}, the performance of pre-trained CNN models significantly drops when adapting to small-scale CXR datasets. In addition, global representations in existing CNNs are formed by simple concatenation and pooling operations. Meanwhile, interactions between local features extracted from different spatial regions and between local and global features are not taken into account. These interactions are not only important to determining the disease from CXR images but also crucial to localising affected regions, consolidating medial diagnoses and treatments.

\subsection{Vision Transformer-based methods}

Vision Transformer (ViT)~\cite{dosovitskiy2020image} is a neural network architecture built upon the encoder of the so-called Transformer~\cite{vaswani2017attention}, that was originally proposed for language translation tasks. When applying to visual domain, ViT decomposes an image into local patches which correspond to tokens in text and natural language processing. These local patches capture local information and can be transformed sequentially multiple times through the architecture. To make ViT an image classifier, a learnable classification token is inserted into the sequence of the local patches. 

ViT has also shown its superiority over traditional CNNs in many medical image analysis tasks~\cite{Shamshad2022TransformersIM,DBLP:journals/corr/abs-2206-01136}. For instance, Matsoukas et al.~\cite{matsoukas2021time} compared the performance of ViT and CNNs in CXR image classification and showed that ViT can capture more complex patterns in CXR images than traditional CNNs. In addition, the authors indicated that using pre-trained weights obtained from ImageNet could further boost up the performance of ViT in CXR image classification. They eventually reached out a conclusion that it is time to switch to ViT for medical image classification. 

In general, the success of ViT lies in several factors. Firstly, this architecture can model both local patterns (captured by local patches) and global correlations between them via attention mechanism. Secondly, although the model is robust and can generalise well to different domains, it is simple in computation yet adaptive and scalable to various problem settings (as the architecture of ViT is mainly made of multi-layer perceptrons). However, existing work focuses only on the representation learning ability of the architecture while overlooking its potential in multi-task learning. Local patterns learnt in ViT are mostly used to support a final goal, e.g., classifying an input image. Meanwhile, these local patterns could be learnt more effectively in a supervised manner via a multi-task learning setting.

\subsection{Multi-task methods}

Multi-task learning has been widely applied in computer vision and image understanding. Examples of the multi-task setting in computer vision include object detection and instance segmentation~\cite{maskrcnn}, semantic and instance segmentation (panoptic segmentation)~\cite{Kendall_CVPR_2018}, anomaly detection and localisation~\cite{huang2022self}. In general, an architecture enabling multi-task learning includes a general backbone for feature learning followed by several heads, each of which handles a sub-task. For instance, Kendall et al.~\cite{Kendall_CVPR_2018} adopted a shared encoder to learn common features and two decoders specialised for each semantic and instance segmentation task. Nevertheless, joint learning of multiple tasks from shared features learnt in early stages of a deep architecture weakly incorporates the sub-tasks. As a consequence, a well-known issue of this method is the multi-task learning may degrade the performance of individual tasks. In~\cite{huang2022self}, the authors applied self-supervised learning on local image regions (masks) for image inpainting in a training set (of normal images). They then used this network on an abnormal image to identify regions of abnormality by identifying high inpainting error regions. In training of the inpainting network, random masks were created and applied to an input image. An autoencoder was then trained to reconstruct both the input image and the mask. At inference phase, masks were initialised as checker board masks at different resolutions. An iterative process was then repeated until convergence where the masks of regions with high reconstruction error were kept while the ones with low error were removed. The final mask was the weighted average and used as the anomaly region prediction for the input image.

Despite existing application of multi-task learning, there are less works in medical image analysis taking the multi-task learning approach, probably due to a lack of multi-labelled medical image datasets in the field. There are methods jointly performing classification of lung related diseases and identification of affected regions from CXR images, e.g.,~\cite{wang2017chestx,yan2018weakly,viniavskyi2020weakly,ouyang2019weakly}. These methods make use of saliency maps (also called attention maps) to define affected regions. However, those attention maps are not learnt explicitly but interpreted additionally to the classification task. Therefore, as shown in experiments, they often include spurious features. For instance, Singla and Feizi~\cite{singla2021salient} showed that an attention map can result in spurious features that are likely to co-occur with the region of interest but not a part of it, e.g., the attribute \textquotedblleft fingers\textquotedblright{} was used for class \textquotedblleft band aid\textquotedblright{} since they usually co-occur. In addition, a common challenge of this approach is that it requires tremendous effort from human annotators for labelling of the attention maps.

Recently, some medical image recognition methods focus on interpreting prediction outcomes. For instance, in~\cite{cao2023robust}, a three-stage network was proposed to jointly perform disease prediction and segmentation. In the first stage, a contrastive learning network was adopted to learn pixel-level spatial consistency. In the second stage, a fracture detection model with ResNeXt~\cite{DBLP:conf/cvpr/XieGDTH17} was built to classify between positive class, with fracture, and negative class, without fracture. In the third stage, a multi-task fracture segmentation model was built to do semantic segmentation and  boundary segmentation. 

Like~\cite{cao2023robust}, Gu et al.~\cite{gu2020vinet} developed a visually interpretable network in place of a traditional CNN to provide not only predictions but also visual hints that lead to predictions. To achieve this the network estimates the importance of every pixel on an input image using an auxiliary network. It then replaces unimportant pixels with random noise, and finally uses the resulting image for predicting an outcome. A mask of important pixels can be used to explain the prediction outcome. Our method also follows this direction, i.e., estimating the importance of image regions on a CXR image. However, we predict patch-level importance instead of pixel-level importance, and use this importance to leverage the estimation of an importance mask from the input CXR image. Our network, in addition, utilises limited amount of patch-level labels to improve the estimation of the importance mask.


\section{Proposed Method}
\label{sec:proposed_method}

\begin{figure*}
\begin{centering}
\includegraphics[width=1.0\textwidth]{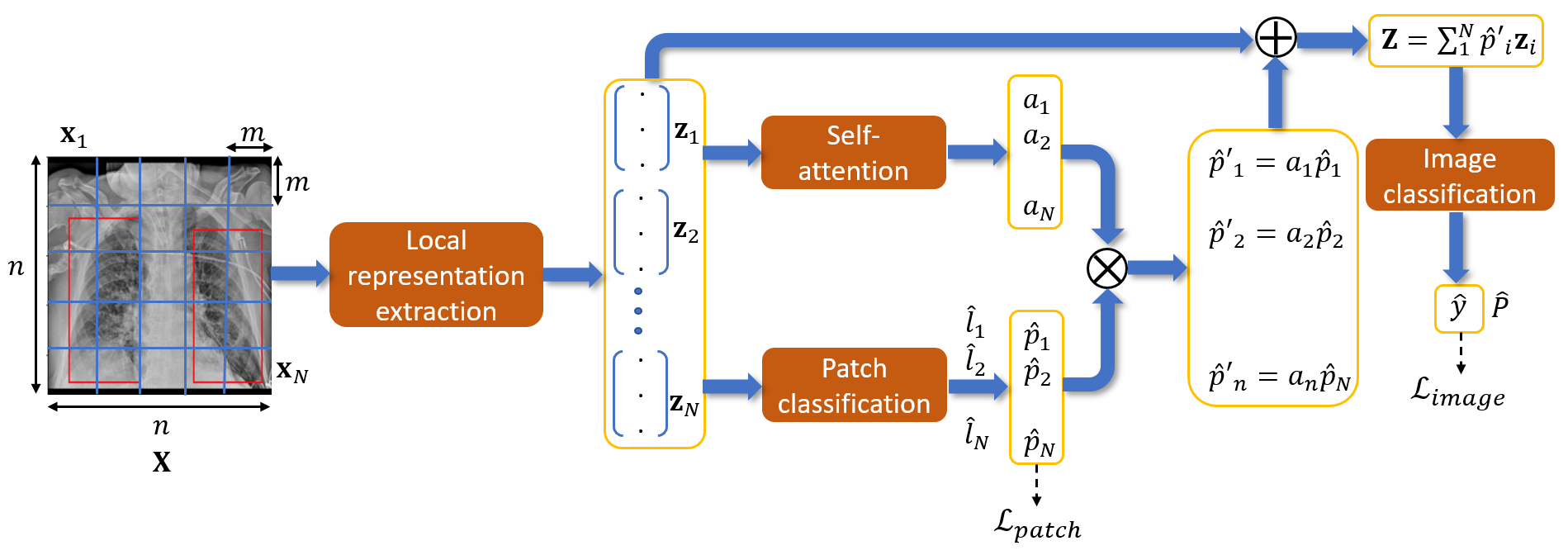}
\end{centering}
\caption{Overview of MVC. Given an input CXR image $\mathbf{X}$, we first pass the image $\mathbf{X}$ to a pre-trained ViT model to obtain local representations for all the local patches on the image. Each local representation $\mathbf{z}_i$ is then fed to a self-attention module to generate a score $a_i$. At the same time, each $\mathbf{z}_i$ is classified by a patch classification module, which is a shared multi-layer perceptron (MLP), to produce a binary label $\hat{l}_i \in \{0,1\}$ and a probability $\hat{p}_i$ that $\mathbf{z}_i$ is an affected region ($\hat{l}_i=1$). Next, we combine the scores $a_i$ and probabilities $\hat{p}_i$ using element-wise multiplication, resulting in a set of weights $\hat{p}'_i = a_i \hat{p}_i$. A global representation $\mathbf{Z}$ is finally created by weighted averaging the local representations $\mathbf{z}_i$, where the weights are $\hat{p}'_i$. This global representation $\mathbf{Z}$ is fed to an image classifier (an MLP) for recognising a disease from the input image.}
\label{fig:MVC}
\end{figure*}

The problem that we solve in this paper is to determine a COVID-19 related disease from a CXR image and, at the same time, to localise affected regions typical for the disease on the input CXR image. In this work, we propose to solve both the sub-tasks simultaneously in a unified framework. 

\subsection{Architecture of MVC}

To address the aforementioned problem, we propose a new architecture, namely multi-task vision transformer for chest X-ray images (MVC). We adopt the ViT model in~\cite{dosovitskiy2020image} as the main backbone for our architecture for several reasons. Firstly, ViT is proven for its capability of learning both local patterns and global correlations between the local patterns. This local-global representation learning well fits our problem as we aim to simultaneously classify a CXR image and identify affected regions (from local patterns). Secondly, ViT is scalable to different problem settings and domains. Thirdly, as shown in the literature and also from our experimental results, ViT outperforms all existing backbones.

Our proposed MVC extends the traditional ViT~\cite{dosovitskiy2020image} in the following aspects. (i) In addition to predicting a lung related disease for an input CXR image as conventional CXR image classification, we also classify local image regions, called patches, on the input CXR image into two classes: positive (affected region) and negative (non-affected region). (ii) We combine learnt representations of local patches with patch classification outcomes in a weighted linear combination where weights are derived from self-attention and patch classification. (iii) We formulate our problem in a multi-task learning setting where sub-tasks include image classification and patch classification. These two sub-tasks are jointly optimised. As shown in our experiments, the tasks support each other, and one task can leverage the other one, leading to improved overall performance. 

The MVC consists of four modules: a local representation extraction module built upon the ViT~\cite{dosovitskiy2020image} to learn representations for local patches, a self-attention module for calculating attention scores from local patches, a patch classification module for classifying local patches, and an image classification module for classifying the entire input image. We illustrate our proposed MVC in Fig.~\ref{fig:MVC}. and describe its main components in corresponding sub-sections. 

\subsubsection{Local representation extraction via ViT}

Let $\mathbf{X}$ be an input CXR image of size $n\times n$ (pixels), and $m\times m$ (pixels) be the resolution of each local patch on the image $\mathbf{X}$. We first apply a uniform grid of size $N=\frac{n}{m} \times \frac{n}{m}$ to $\mathbf{X}$. We choose $n$ and $m$ such that $n/m$ is integer for computational convenience. This operation results in a sequence of local patches $(\mathbf{x}_{1},\dots,\mathbf{x}_{N})$ ordered from left-to-right top-to-bottom, where $\mathbf{x}_i \in \mathbb{R}^{m \times m}$, $i=1,\dots,N$. We then apply the pre-trained ViT model in~\cite{dosovitskiy2020image} to the local patches $(\mathbf{x}_{1},\dots,\mathbf{x}_{N})$ to obtain local representations $(\mathbf{z}_i,\dots,\mathbf{z}_{N})$ as,
\begin{align}
(\mathbf{z}_{1},\dots,\mathbf{z}_{N})=\text{ViT}^{m}(\mathbf{x}_{1},\dots,\mathbf{x}_{N})
\label{eq:ViT-feature-extraction}
\end{align}
where $\text{ViT}^{m}$ denotes the ViT model with $m \times m$-pixel patches. 

Note that, for local representation extraction, we only use patch embedding and attention layers except for prediction heads from the original ViT model. In general, this step transforms a local image patch $\mathbf{x}_i \in \mathbb{R}^{m \times m}$ into an embedding vector $\mathbf{z}_i \in \mathbb{R}^{d}$ which is more informative about the importance of the patch $\mathbf{x}_i$ in relation to other patches.

\subsubsection{Patch classification}

Local representations $\mathbf{z}_i$ encoded by the ViT model are classified by a patch classifier into two classes: positive or negative. A positive label indicates a disease related region (affected region) while a negative label implies a normal region (non-affected region). We realise the patch classifier with a 2-layer MLP with ReLU activation function and a softmax layer for the output layer. The MLP takes input as a vector $\mathbf{z}_i$ and returns a label $\hat{l}_i \in \{0,1\}$ with a probability $\hat{p}_i \in [0,1]$ that $\mathbf{x}_i$ is a positive (affected) patch,
\begin{align}
(\hat{l}_i,\hat{p}_{i}) =\text{Patch-Classifier}(\mathbf{z}_{i})
\label{eq:patch-classification}
\end{align}

\subsubsection{Self-attention}

Local representations $\mathbf{z}_i$ are also passed to a self-attention module to calculate a set of attention scores $a_i$. Specifically, each representation $\mathbf{z}_i$ is fed to a 2-layer MLP with ReLU activation function, then normalised by a softmax function to produce a score $a_i$,
\begin{align}
a_{i} = \frac{\exp(\mathbf{w}_i^\top \text{ReLU}(\mathbf{V} \mathbf{z}_i))}{\sum_{j=1}^{N} \exp(\mathbf{w}_j^\top \text{ReLU}(\mathbf{V} \mathbf{z}_j))}, i=1,\dots,N
\label{eq:softmax}
\end{align}
where we use a separate parameter vector $\mathbf{w}_i \in \mathbb{R}^{h}$ for each representation $\mathbf{z}_i$ and a shared parameter matrix $\mathbf{V} \in \mathbb{R}^{h \times d}$ for all the representations; $d$ is the dimension of $\mathbf{z}_i$ and $h$ is the size of the hidden layer in the MLP (see the second row in Table~\ref{tab:MLPs}).

\subsubsection{Image classification}

Attention scores $a_i$ achieved from the self-attention module and positive probabilities $\hat{p}_i$ calculated from the patch classification module are then combined via an element-wise multiplication operator to produce a set of attention-based positive scores $(\hat{p}'_1,\dots,\hat{p}'_{N})$,
\begin{align}
(\hat{p}'_1,\dots,\hat{p}'_{N}) = (a_1\hat{p}_1,\dots,a_N \hat{p}_{N})
\label{eq:attention-based-positive-probabilities}
\end{align}

This combination aims to regulate the prediction of each local patch by the attention score of the patch. For instance, the attention score of a false positive patch can mitigate the prediction probability of that patch, affecting to the final decision to a disease on the entire input image. The scores $\hat{p}'_i$ are then used to weight the local representations $z_i$ to create a global representation $\mathbf{Z} \in \mathbb{R}^d$ as,
\begin{align}
\mathbf{Z} =\sum_{i=1}^{N}\hat{p}_{i}' \mathbf{z}_{i}
\label{eq:global-representation}
\end{align}

The global representation $\mathbf{Z}$ is finally fed to an image classifier, which is another 2-layer MLP with ReLU activation function and a softmax layer for the last layer, to identify a lung related disease $\hat{y}$ and its probability $\hat{P}(\hat{y})$, 
\begin{align}
(\hat{y},\hat{P}(\hat{y})) = \text{Image-Classifier}(\mathbf{Z})
\label{eq:global-classification}
\end{align}
where $\hat{y}$ is from a set of predefined disease classes $Y$, and $\hat{P}$ is a probability distribution of $\hat{y}$ achieved from the softmax layer, i.e., $\hat{y}=\arg\max_{\bar{y} \in Y} \hat{P}(\bar{y})$. 

\subsection{Multi-task learning}
\label{sec:multi-task-learning}

The proposed MVC can be trained end-to-end, and sub-tasks (i.e., patch classification and image classification) can be jointly optimised during training. Specifically, suppose that each training CXR image $\mathbf{X}$ is associated with a label $y \in Y$ and a sequence of true labels $\mathbf{l}=(l_{1},\dots,l_N) \in \{0,1\}^{N}$ for the local patches $(\mathbf{x}_{1},\dots,\mathbf{x}_N)$. 

The label set $Y$ includes lung related diseases, and is predefined. We obtain the patch labels $l_{1},\dots,l_N$ as follows. Suppose that affected regions of an image $\mathbf{X}$ are delineated as bounding boxes and provided in the ground-truth. A patch $\mathbf{x}_{i}$ is determined as positive ($l_{i}=1$), if it matches an affect region in the ground-truth, and negative ($l_{i}=0$), otherwise. A match is confirmed if there exists an affected region $R$ in the ground-truth data such that $\frac{|\mathbf{x}_{i} \cap R|}{|\mathbf{x}_{i}|} > 0.5$, where $\cap$ denotes the intersection of two regions and $|\cdot|$ represents the area of a region (in number of pixels).

\begin{figure}
\includegraphics[width=0.49\columnwidth]{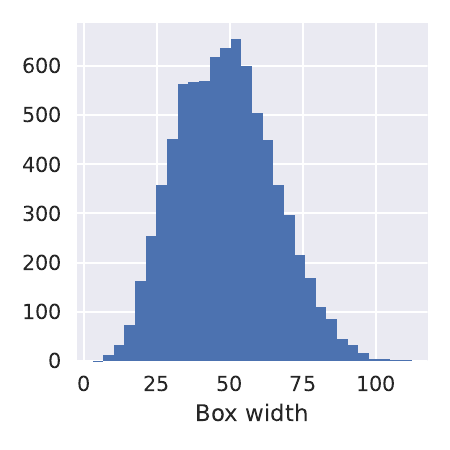}
\includegraphics[width=0.49\columnwidth]{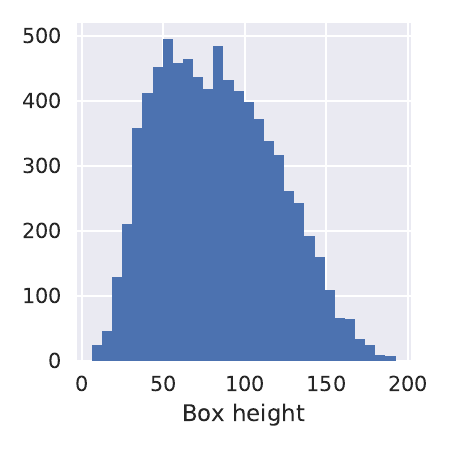}
\caption{Distributions of the width and height of affected regions' bounding boxes. The y-axis shows counts.}
\label{fig:width-height-histograms}
\end{figure}

Note that, we do not formulate the task of finding affected regions as an object detection problem. This is because, although affected regions are provided as bounding boxes in the ground-truth, they do not really represent meaningful objects as in the object detection setting. Instead, those regions vary in their shape and size, and can be scattered in a CXR image. In our method, we design the local patches such that they can capture enough information to determine a disease while being able to represent smallest affected regions. Specifically, we set the size of the local patches to $16 \times 16$-pixels in relation to a resolution of $224 \times 224$-pixels. This size is estimated empirically from the minimum of the dimensions of affected regions' bounding boxes from the ground-truth. Fig.~\ref{fig:width-height-histograms} shows the distributions of the width and height of affected regions' bounding boxes.

In Eq.~(\ref{eq:global-classification}), a predicted class label $\hat{y}$ for the input image $\mathbf{X}$ is obtained via a softmax layer. Let $P$ be the target probability distribution of true labels $y$, i.e., $P(\bar{y})=1$ if $\bar{y}=y$, and $P(\bar{y})=0$, otherwise. We define a classification loss $\mathcal{L}_{image}$ for the image classification task using cross-entropy loss as,
\begin{align}
    \mathcal{L}_{image}(y, \hat{y}) = -\sum_{\bar{y} \in Y} P(\bar{y}) \log \hat{P}(\bar{y}) = -\log \hat{P}(y)
    \label{eq:image-loss}
\end{align}

From Eq.~(\ref{eq:patch-classification}), we construct $\mathbf{\hat{l}}=(\hat{l}_{1},\dots,\hat{l}_N)$ and $\mathbf{\hat{p}}=(\hat{p}_{1},\dots,\hat{p}_N)$. Let $\mathbf{p}=(p_{1},\dots,p_N)$ be the sequence of probabilities of the sequence of true labels $\mathbf{l}=(l_1,\dots,l_N)$ being positive, i.e., $p_i=1$ if $l_i=1$ and $p_i=0$, otherwise. We define a loss $\mathcal{L}_{patch}$ for the patch classification module using binary cross-entropy loss as,
\begin{align}
    \mathcal{L}_{patch}(\mathbf{l}, \mathbf{\hat{l}}) = -\sum_{i=1}^{N} p_i \log \hat{p}_i + (1- p_i) \log (1 - \hat{p}_i)
    \label{eq:patch-loss}
\end{align}

Finally, we define a loss $\mathcal{L}$ to train the entire MVC as,
\begin{align}
    \mathcal{L} = \mathcal{L}_{image}(y,\hat{y}) + \frac{1}{N} \mathcal{L}_{patch}(\mathbf{l}, \mathbf{\hat{l}})
    \label{eq:general-loss}
\end{align}

\section{Experiments}
\label{sec:exp}

\subsection{Experimental setup}

\subsubsection{Dataset}

Since we aim to both classify CXR images and identify affected regions, we chose the COVID-19 Chest-X-ray dataset in~\cite{vaya2020bimcv}, which is associated with both image-level and region-level labels, to conduct experiments. We note that the labels in the COVID-19 Chest-X-ray dataset are reliable as they are provided by medical experts. In addition, COVID-19 Chest-X-ray is the largest public COVID-19 image dataset in the field.

This dataset consists of 5,937 images, categorised into 4 classes: Typical, Atypical, Indeterminate, and Negative. The original CXR images in the dataset are in $256\times256$-pixel resolution. We first resized all the images to $224\times224$-pixel resolution. The dataset is split into a training set including 4,749 images and a test set including 1,188 images. We describe the COVID-19 Chest-X-ray dataset in Table~\ref{tab:dataset}. We also present the average ratio of affected regions (in pixels) and their entire image on classes labelled with affected regions in Table~\ref{tab:dataset}.

\subsubsection{Implementation details}

We initialised the local representation extraction module from a pre-trained ViT model~\cite{dosovitskiy2020image}, trained on ImageNet~\cite{imagenet_cvpr09}. Recall that, the local representation extraction module is constructed from all the layers in the ViT except for the last layer.

Our architecture makes use of several MLPs in the sub-modules (patch classification, self-attention scores calculation, and image classification). We describe the details of these MLPs in Table~\ref{tab:MLPs}. Note that, ReLU activation function is used in all the MLPs.

We trained the MVC using 15 epochs, learning rate of 0.0001, batch size of 16, and ADAM optimiser. We also applied data augmentation to the training of our model and other baselines for fair comparisons. The data augmentation includes horizontal flipping, affine transformations, and colour jittering. The augmentation was performed randomly on training batches. This helps to prevent overfitting in the training and makes the model more robust against noise and variations such as image misalignment, image scaling, making bounding boxes not fitting well into affected regions.

We implemented our proposed MVC and other methods in Pytorch 1.10~\cite{pytorch2019} and conducted all experiments on 2 NVIDIA GeForce RTX 2080 Ti. 

\begin{table}

\caption{Details of the COVID-19 Chest-X-ray dataset~\cite{vaya2020bimcv}.}

\begin{centering}
\begin{tabular}{l|c|c|c|c} \toprule
               & \multicolumn{3}{c|}{\textbf{No. images}} & \textbf{\% Affected regions} \\ \cmidrule{2-4}
\textbf{Class} & \textbf{Training} & \textbf{Testing} &                     \textbf{Total} & \textbf{per image} \\ \midrule
Negative      & 1,317 & 330 & 1,647 & - \\
Typical       & 2,280 & 570 & 2,850 & 27\% \\
Atypical      & 313 & 78 & 391 & 10\% \\
Indeterminate & 839 & 210 & 1,049 & 12\% \\ \midrule
All           & 4,749 & 1,188 & 5,937 & - \\ \bottomrule
\end{tabular}
\par\end{centering}

\label{tab:dataset}
\end{table}

\begin{table}

\caption{Details of the MLPs used in our MVC. The input size of all the MLPs is 384, which is the dimension $d$ of local representations $\mathbf{z}_i$. The output size of the MLP used in self-attention is 1 as this MLP produces an attention score for each local patch. For the MLPs used in patch classification and image classification, the output size is the number of classes, e.g., 2 classes (binary output) for patch classification and 4 classes (4 lung related diseases) for image classification.}

\begin{centering}
\begin{tabular}{l|c|c|c} \toprule
\textbf{MLP} & \textbf{Input size} & \textbf{Hidden size} & \textbf{Output size} \\ \midrule
Patch classification & 384 & 500 & 2 \\
Self-attention & 384 & 500 & 1 \\
Image classification & 384 & 500 & 4 \\
\bottomrule
\end{tabular}
\par\end{centering}

\label{tab:MLPs}
\end{table}

\subsection{Result analysis}

\subsubsection{CXR image classification}

\begin{table*}
\caption{Comparison of our proposed MVC with existing baselines regarding to the number of parameters (in millions), model size (in mega bytes), floating point operations - FLOPs (in millions), and recognition accuracy (in \%).} 
\begin{centering}
\begin{tabular}{l|c|c|c|c|c|c|c|c} \toprule
& \multicolumn{3}{c|}{\textbf{Specification}} &  \multicolumn{5}{c}{\textbf{Recognition accuracy (\%)}} \\ \cmidrule{2-9}
\textbf{Model} & \textbf{No. Params (M)} & \textbf{Size (MB)} & \textbf{FLOPs (M)} & \textbf{Negative} & \textbf{Typical} & \textbf{Atypical} & \textbf{Indeterminate} & \textbf{Overall} \\ \midrule
AlexNet~\cite{DBLP:conf/nips/KrizhevskySH12} & 57 & 217 & 711 & 77.0 & 54.2 & 28.2 & 29.1 & 54.4 \\
DenseNet~\cite{huang2017densely} (used in ~\cite{rajpurkar2017chexnet}) & 7 & 27 & 2881 & 72.7 & 66.8 & 32.1 & 21.0 & 58.1 \\
ResNet-50~\cite{he2016deep} & 11 & 43 & 1821 & 74.9 & 64.7 & 18.0 & 29.5 & 58.3 \\
SqueezeNet~\cite{DBLP:journals/corr/IandolaMAHDK16} & 1 & 3 & 269 & 52.7 & 70.7 & 30.8 & 30.5 & 58.4 \\
VGG~\cite{simonyan2014very} (used in~\cite{apostolopoulos2020COVID}) & 129 & 491 & 7641 & 82.4 & 65.1 & 14.1 & 22.4 & 59.1 \\
$\text{ViT}^{8}$~\cite{dosovitskiy2020image} & 22 & 83 & 87 & 70.3 & 70.2 & 20.5 & 23.8 & 58.8 \\
$\text{ViT}^{16}$~\cite{dosovitskiy2020image} & 22 & 83 & 60 & 86.7 & 67.7 & 28.2 & 13.3 & 60.7 \\ \midrule
\textbf{Our MVC} & 22 & 84 & 60 & 84.5 & 63.2 & 28.2 & 33.8 & 61.6 \\ \bottomrule
\end{tabular}
\par\end{centering}
\label{tab:image-classification}
\end{table*}

\begin{figure}
\centering
\includegraphics[width=0.7\columnwidth]{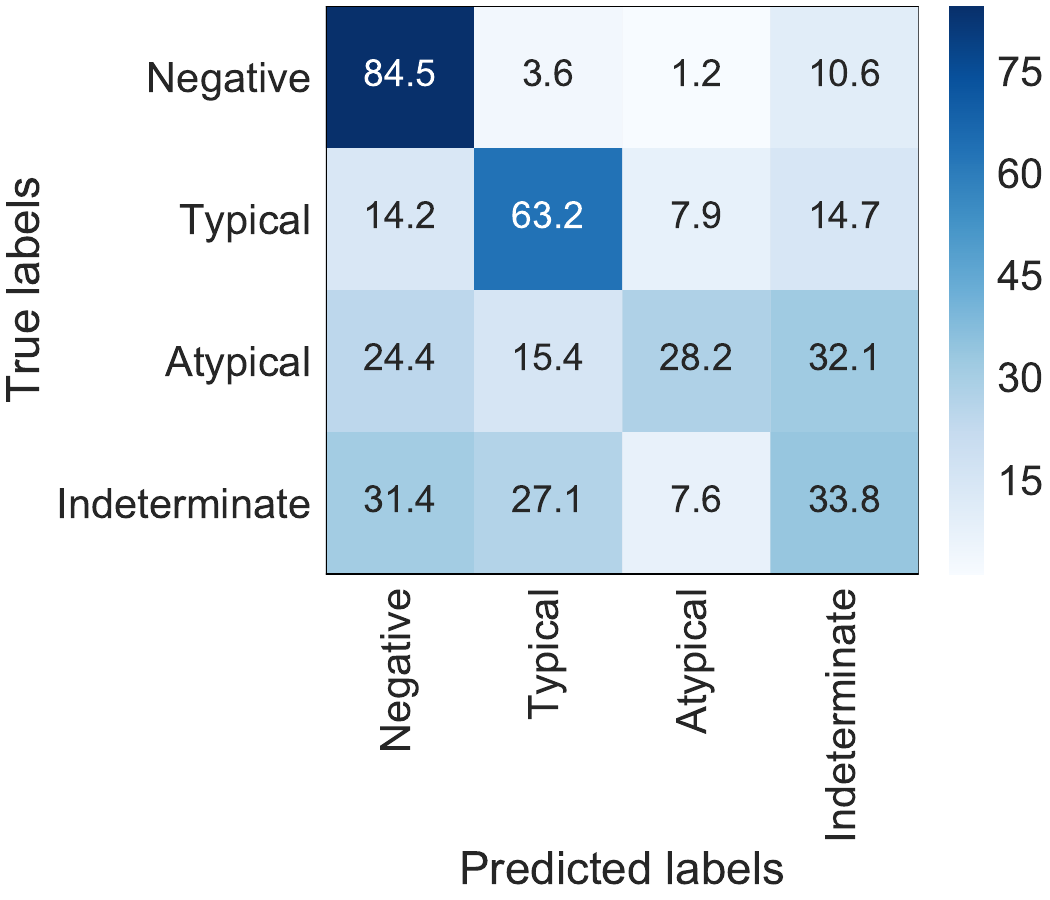}
\caption{Confusion matrix of our MVC on the test set of the COVID-19 Chest-X-ray dataset.}
\label{fig:confusion-matrix}
\end{figure}

We evaluated our method in the CXR image classification task and measure its performance via the recognition accuracy on every disease class and overall. We provide a confusion matrix showing the recognition accuracy of our method on the test set of the COVID-19 Chest-X-ray dataset in Fig.~\ref{fig:confusion-matrix}.

We also compared our proposed method with prevailing image classification baselines including AlexNet~\cite{DBLP:conf/nips/KrizhevskySH12}, DenseNet~\cite{huang2017densely} (used in ~\cite{rajpurkar2017chexnet}), ResNet-50~\cite{he2016deep} (used in~\cite{narin2021automatic}), SqueezeNet~\cite{DBLP:journals/corr/IandolaMAHDK16},   VGG~\cite{simonyan2014very} (used in~\cite{apostolopoulos2020COVID}), Vision Transformer~\cite{dosovitskiy2020image}
at resolutions $8\times8$ ($\text{ViT}^8$) and $16\times16$ ($\text{ViT}^{16}$) for local patches. For the baselines, we adopted pre-trained models and customised them by replacing the last layer to fit with the disease classes in our problem. We then fine-tuned the baselines on the COVID-19 Chest-X-ray dataset.

We summarise the specification of all the baselines and their recognition accuracy (on every class and overall) in the image classification task in Table~\ref{tab:image-classification}. As shown in the results, compared with the existing baselines, our proposed MVC achieves the best overall performance. $\text{ViT}^{16}$ ranks second and surpasses traditional CNNs. Our MVC also outperforms $\text{ViT}^{16}$, which is used as the backbone for local representation extraction. This is evident for advantages brought by the incorporation of local and global information, and by patch classification to image classification.

Except for SqueezNet, all other baselines (including the MVC) achieve the best and second best accuracy on the Negative and Typical class respectively. Atypical and Indeterminate classes remain challenging to all the methods, probably due to limited training data.

In this experiment, we also compared two variants of the $\text{ViT}$: $\text{ViT}^{8}$ and $\text{ViT}^{16}$ corresponding to $(8 \times 8)$- and $(16 \times 16)$-pixel local patch setting. We observed that $\text{ViT}^{16}$ outperforms $\text{ViT}^{8}$ in the image classification task. Hence, we adopted $\text{ViT}^{16}$ as the backbone for local representation extraction in our proposed MVC.

\begin{table}
\caption{Performance of our proposed MVC in affected region identification using various metrics and on different disease classes. Higher values indicate better performances.} 
\begin{centering}
\begin{tabular}{l|c|c|c|c} \toprule
               & \multicolumn{4}{c}{\textbf{Performance metric}}\\
               \cmidrule{2-5}
\textbf{Class} & \textbf{F1-score} & \textbf{AU-ROC} & \textbf{AU-PR} & \textbf{Jaccard} \\ \midrule
Typical & 68.5 & 95.3 & 76.2 & 42.8 \\
Atypical & 38.7 & 94.5 & 35.0 & 8.6 \\
Indeterminate & 49.3 & 95.0 & 49.9 & 17.9 \\
Overall & 65.8 & 94.8 & 72.8 & 42.8 \\ \bottomrule
\end{tabular}
\par\end{centering}
\label{tab:patch-classification}
\end{table}

\subsubsection{Affected region identification}

Since we formulate the task of affected region identification as binary classification of local patches, we measured the performance of our method in this task using common metrics in binary classification including F1-score, area under receiver operating characteristic curve (AU-ROC), and area under precision-recall curve (AU-PR). Those metrics reflect the trade-off between true positive and false positive rates in classification of local patches. Recall that the label (affected vs non-affected) for each patch is determined based on the overlapping between the patch and an affected region's bounding box in the ground-truth. We also measured the Jaccard similarity between all affected regions on a CXR image identified by our method and regions' bounding boxes given in the ground-truth. Different from F1-score, AU-ROC, and AU-PR, the Jaccard similarity measures the coincidence of predicted regions and ground-truth regions at pixel level as the Jaccard similarity is calculated on region masks generated from affected patches predicted by our method and regions' bounding boxes from the ground-truth. 

\begin{figure*}
    \begin{centering}
    \includegraphics[width=3.0cm]{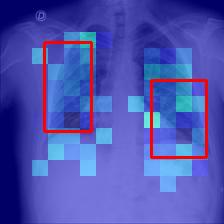} 
    \includegraphics[width=3.0cm]{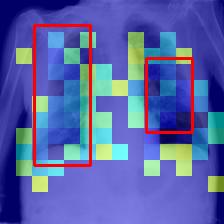} 
    \includegraphics[width=3.0cm]{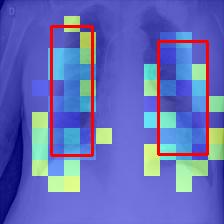} 
    \includegraphics[width=3.0cm]{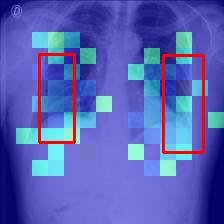} 
    \includegraphics[width=3.0cm]{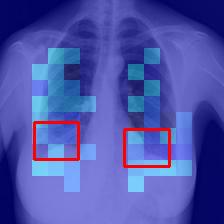} \\
    \text{Our MVC} \\
    \vspace{0.3cm}
    \includegraphics[width=3.0cm]{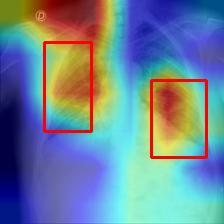} 
    \includegraphics[width=3.0cm]{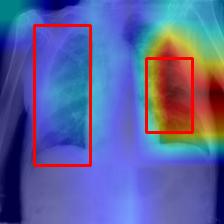} 
    \includegraphics[width=3.0cm]{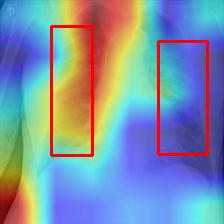} 
    \includegraphics[width=3.0cm]{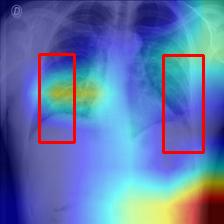} 
    \includegraphics[width=3.0cm]{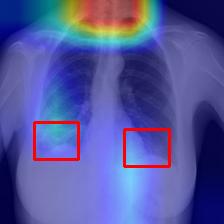}\\
    \text{AlexNet~\cite{DBLP:conf/nips/KrizhevskySH12}} \\
    \vspace{0.3cm}
    
    \includegraphics[width=3.0cm]{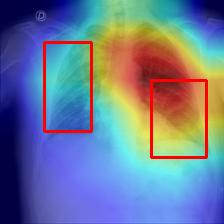} 
    \includegraphics[width=3.0cm]{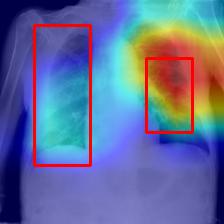} 
    \includegraphics[width=3.0cm]{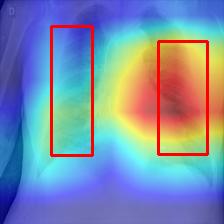} 
    \includegraphics[width=3.0cm]{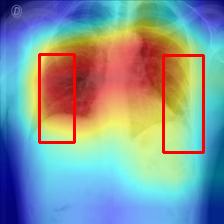} 
    \includegraphics[width=3.0cm]{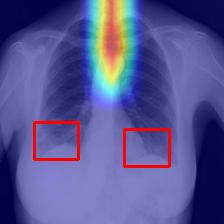}\\
    \text{DenseNet~\cite{huang2017densely}} \\
    \vspace{0.3cm}
    
    \includegraphics[width=3.0cm]{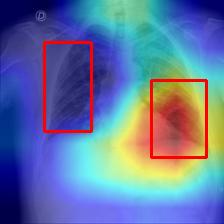} 
    \includegraphics[width=3.0cm]{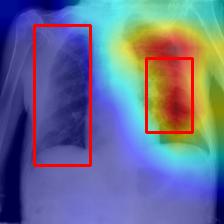} 
    \includegraphics[width=3.0cm]{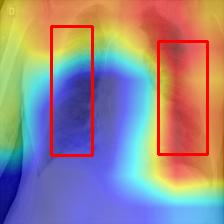} 
    \includegraphics[width=3.0cm]{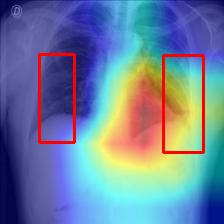} 
    \includegraphics[width=3.0cm]{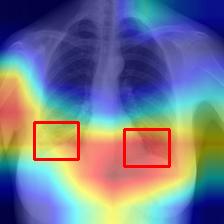}\\
    \text{ResNet-50~\cite{he2016deep}} \\
    \vspace{0.3cm}
    
    \includegraphics[width=3.0cm]{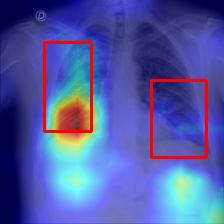} 
    \includegraphics[width=3.0cm]{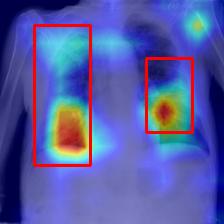} 
    \includegraphics[width=3.0cm]{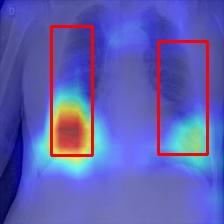} 
    \includegraphics[width=3.0cm]{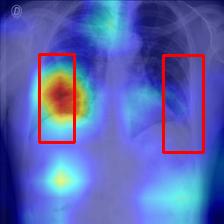} 
    \includegraphics[width=3.0cm]{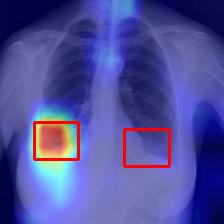}\\
    \text{SqueezeNet~\cite{DBLP:journals/corr/IandolaMAHDK16}} \\
    \vspace{0.3cm}
    
    \includegraphics[width=3.0cm]{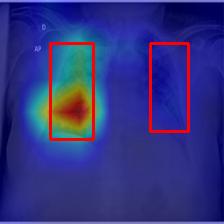} 
    \includegraphics[width=3.0cm]{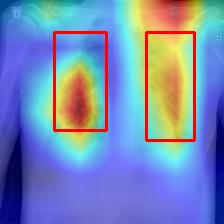} 
    \includegraphics[width=3.0cm]{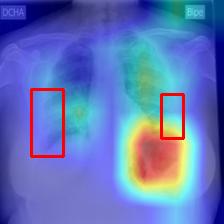} 
    \includegraphics[width=3.0cm]{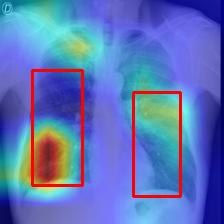} 
    \includegraphics[width=3.0cm]{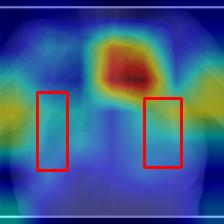}\\
    \text{VGG~\cite{simonyan2014very}} 
    
    \caption{Results of affected regions identified by our method (marked by patches classified as positive) and by other methods (shown in Grad-CAM heat maps~\cite{selvaraju2017grad}). Ground-truth regions are highlighted (in red boxes).} 
    \label{fig:visual-results}
    \end{centering}
\end{figure*}

We report F1-score, AU-ROC, AU-PR, and the Jaccard similarity of our MVC on every class and overall in Table~\ref{tab:patch-classification}. We illustrate several region identification results of our method in Fig.~\ref{fig:visual-results} (the first row). As shown in the results (square patches), our method can well identify affected regions. Moreover, our identified regions are even better localised in lungs areas, compared with the ground-truth data (red boxes). 


Fig.~\ref{fig:visual-results} also visually compare our method with existing baselines in affected region identification. Since our method is the only method explicitly aiming to identify affected regions in addition to disease classification, we showcase the ability of affected region identification from existing methods via heat maps. Particularly, we applied the Gradient-weighted Class Activation Mapping (Grad-CAM) in~\cite{selvaraju2017grad} to extract the heat maps. For each model, we used the gradients of positive scores from the final convolutional layer/feature map in the model to produce a course heat map. The heat maps highlight important regions from an image which make the most influence to the final classification outcome (a disease). As shown in Fig.~\ref{fig:visual-results}, affected regions are not clearly indicated in the heat maps of existing methods. AlexNet, DenseNet, and ResNet tend to use information from regions spreading a large proportion of the input image whereas positive regions only are accounted for a small proportion. SqueezeNet and VGG on the other hand are more localised but their identified important regions do not well cover real affected regions. Therefore, to make predicted regions more accurately, a post-processing step is required to the baseline methods.

\subsubsection{Ablation study}
\label{sec:ablation-study}

We investigated different settings in the design of our MVC. In particular, we compared two variants of the MVC built upon two patch size setting, $8\times8$-pixels ($\text{ViT}^8$) and $16\times16$-pixels ($\text{ViT}^{16}$) for local patches. Table~\ref{tab:ViT8-vs-ViT16} shows the performances of the MVC with $\text{ViT}^{16}$ and $\text{ViT}^{8}$ backbones in both the image classification and affected region identification tasks. We observed that the MVC built with $\text{ViT}^{16}$ consistently outperforms that built with $\text{ViT}^8$ on all the performance metrics. Recall that $\text{ViT}^{16}$ also surpasses $\text{ViT}^{8}$ on the overall image recognition accuracy (see Table~\ref{tab:image-classification}). 

\begin{table}
\caption{Comparison of two variants of MVC with different patch-size settings. Only overall performances are presented.} 
\begin{centering}
\begin{tabular}{l|c|c|c|c|c} \toprule
               & \multicolumn{5}{c}{\textbf{Performance metric}}\\
               \cmidrule{2-6}
\textbf{Model} & \textbf{Acc} & \textbf{F1-score} & \textbf{AU-ROC} & \textbf{AU-PR} & \textbf{Jaccard} \\ \midrule
MVC-$\text{ViT}^8$    & 60.5 & 64.2 & 94.3 & 70.3 & 41.9 \\
MVC-$\text{ViT}^{16}$ & 61.6 & 65.8 & 94.8 & 72.8 & 42.8 \\ \bottomrule
\end{tabular}
\par\end{centering}
\label{tab:ViT8-vs-ViT16}
\end{table}

The novelty of our proposed MVC relies on the joint image classification and affected region identification, leading to improved performance in both the tasks. To validate this capability, we created two variants of the MVC as follows. In the first variant, we skipped the patch classification module and simply set $\hat{l}_i = 1$ for all local patches $\mathbf{x}_i$. This leads to $\mathbf{Z}=\sum_{i=1}^{N}a_i \mathbf{z}_i$. We call this variant ``MVC-image'' as it aims to perform image classification solely. In the second variant, we only trained the patch classification module while freezing both the self-attention and image classification modules. This variant is referred to as ``MVC-patch''. We compared these two variants with the full version of the MVC in Table~\ref{tab:global-local}. Since ``MVC-image'' is not designed for patch classification, region identification metrics are not measured in this variant. Similarly, recognition accuracy is not applied to ``MVC-patch''.

\begin{table}
\caption{Comparison of the full version (multi-task) of MVC with its variants on single-task settings. Only overall performances are presented.}
\begin{centering}
\begin{tabular}{l|c|c|c|c|c} \toprule
               & \multicolumn{5}{c}{\textbf{Performance metric}}\\
               \cmidrule{2-6}
\textbf{Variant} & \textbf{Acc} & \textbf{F1-score} & \textbf{AU-ROC} & \textbf{AU-PR} & \textbf{Jaccard} \\ \midrule
MVC-image & 60.6 & - & - & - & - \\
MVC-patch & - & 65.7 & 94.6 & 71.5 & 42.4 \\
MVC-full & 61.6 & 65.8 & 94.8 & 72.8 & 42.8 \\ \bottomrule
\end{tabular}
\par\end{centering} 
\label{tab:global-local}
\end{table}

We observed that pre-trained models (e.g., the ViT model pre-trained on ImageNet~\cite{imagenet_cvpr09}) also bring benefits to construction of the MVC despite of domain shift. This observation is consistent with findings indicated in other studies~\cite{litjens2017survey,ccalli2021deep}. In this ablation study, we compared the performances of our MVC with and without using pre-trained models (i.e., training from scratch) in both image classification and affected region identification. We report results of this comparison in Table~\ref{tab:pre-trained-models}.

\begin{table}
\caption{Performance of our proposed MVC with and without using pre-trained models. Only overall performances are presented.} 
\begin{centering}
\begin{tabular}{l|c|c|c|c|c} \toprule
               & \multicolumn{5}{c}{\textbf{Performance metric}}\\
               \cmidrule{2-6}
\textbf{Variant} & \textbf{Acc} & \textbf{F1-score} & \textbf{AU-ROC} & \textbf{AU-PR} & \textbf{Jaccard} \\ \midrule
With pre-trained & 61.6 & 65.8 & 94.8 & 72.8 & 42.8 \\
W/o pre-trained & 53.1 & 51.9 & 88.9 & 51.0 & 31.3 \\ \bottomrule
\end{tabular}
\par\end{centering}
\label{tab:pre-trained-models}
\end{table}

\begin{figure*}
\begin{centering}
\subfloat[]{\begin{centering}
\includegraphics[width=0.67\columnwidth]{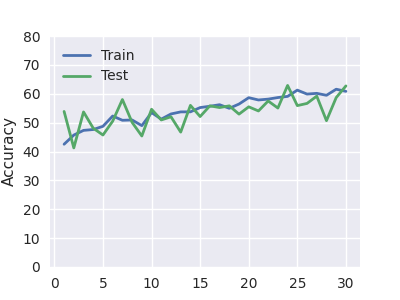}
\par\end{centering}
}
\subfloat[]{\begin{centering}
\includegraphics[width=0.67\columnwidth]{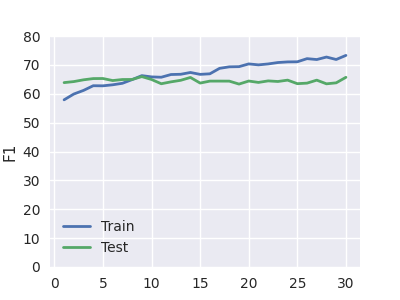}
\par\end{centering}
}
\subfloat[]{\begin{centering}
\includegraphics[width=0.67\columnwidth]{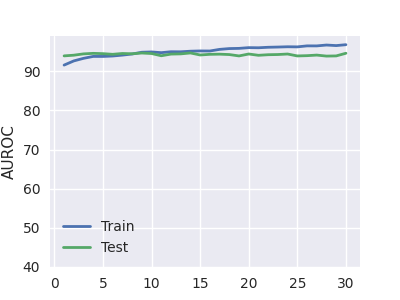}
\par\end{centering}
} \\
\subfloat[]{\begin{centering}
\includegraphics[width=0.67\columnwidth]{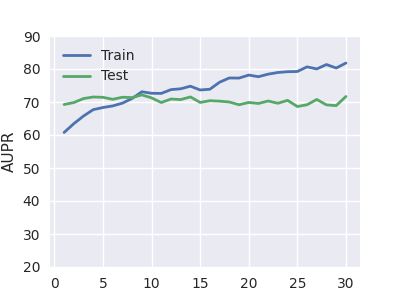}
\par\end{centering}
}
\subfloat[]{\begin{centering}
\includegraphics[width=0.67\columnwidth]{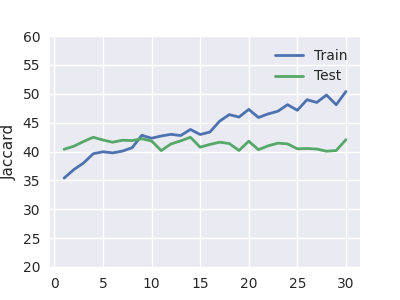}
\par\end{centering}
}
\par\end{centering}
\caption{Learning in the MVC model on the training and test set using different metrics and various epochs. (a) Overall accuracy, (b) F1-score, (c) AU-ROC. (d) AU-PR. (e) Jaccard similarity}
\label{fig:learning-curves}
\end{figure*}

We investigated the learning in the MVC in both image classification and affected region identification using different metrics and under various epochs in Fig.~\ref{fig:learning-curves}. It is shown that the model starts converting at 15 epochs and its performance (on all metrics) on the test set saturates from that point. In our experiments, we stopped the training at 15 epochs even though the model keeps slightly improving on the training set. This early stopping helps the model avoid being overfitted.

\section{Discussion and Conclusion}
\label{sec:discussion}

\subsection{Contributions}

This paper proposes a multi-task vision transformer network, named MVC, for COVID-19 disease recognition and affected region identification from chest x-ray images. Technically, we design our network in a two-task setting: image recognition and sub-region classification, and make the two tasks dually related. To enable such ability, we adopt the backbone of Vision Transformer for learning of local representations. These local representations are used to identify affected regions and combined via self-attention into global representations for disease recognition. One advantage of using local features is their power in describing regions of complex shapes without adding extra effort. Self-attention is also useful to encode the contextual information of the local structures. The entire network can be trained end-to-end and performs both disease recognition and affected region identification simultaneously. These sub-tasks support each other, leveraging the overall performance. To the best our knowledge, such a chest X-ray-based COVID-19 diagnosis approach is novel. 

\subsection{Findings}

The proposed MVC was thoroughly evaluated and compared with existing baselines on a benchmark COVID-19 chest-x-ray dataset to recognise 4 types of lung related diseases including Negative, Typical, Atypical, and Indeterminate. 

The MVC achieved an overall accuracy of 61.5\% in the task of disease recognition. The model performed well on Negative and Typical images. Atypical and Indeterminate appear as challenging classes; they have less training data, compared with other classes. Misclassified cases within the Atypical and Indeterminate classes distribute uniformly across all the classes, meaning the uncertainty of the model to those classes. We also found that, there are much fewer positive (affected) regions in the Atypical and Indeterminate classes, compared with those in the Typical class. Overall, the proposed MVC outperforms existing chest x-ray image classification baselines. 

In addition to recognising diseases from chest x-ray images, the MVC can also explicitly locate affected regions relevant to a classified disease. This is another advantage of our method and would benefit for doctors and radiologists in consolidating their diagnoses. 

Experimental results also confirm the advantage of multi-task learning over its single-task counterparts. Despite improved overall accuracy, the proposed MVC does not increase much computational overhead, additionally to its baseline architecture (the Vision Transformer~\cite{dosovitskiy2020image}).



\subsection{Future work}


The MVC is trained using supervised learning for both image recognition and patch classification. This requires data labelling at both image and region level, limiting the applicability of the method. Since the MVC supports both local and global representation learning, it fits well the formulation of multiple-instance learning where local patches are considered as ``instances'' and images are treated as ``bags''~\cite{DBLP:conf/cvpr/SultaniCS18}. This formulation can relax the requirement of patch labelling and is considered as our future work.


\bibliography{bib.bib}
\bibliographystyle{ieeetr}

\begin{IEEEbiography}
[{\includegraphics[width=1in,height=1.25in,clip,keepaspectratio]{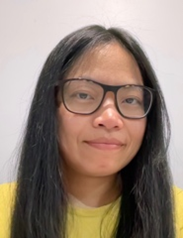}}]{Huyen Tran} received a B.S in Informatics Education from Ho Chi Minh University of Education, Vietnam, and a M.S. in Computer Science from Canberra University, Victorial, Australia. She is currently pursuing her Ph.D. in Computer Science at the School of Information Technology, Deakin University, Australia. Tran's research interests are computer vision and machine learning for medical image analysis.
\end{IEEEbiography}

\begin{IEEEbiography}
[{\includegraphics[width=1in,height=1.25in,clip,keepaspectratio]{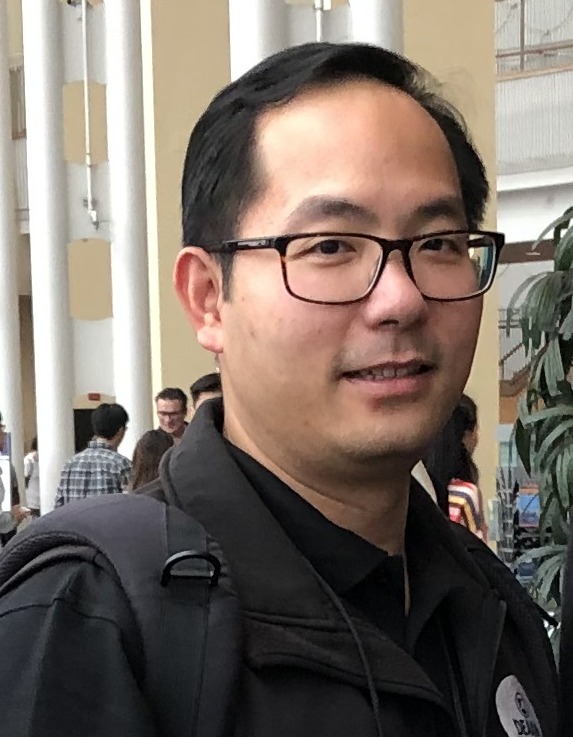}}]{Duc Thanh Nguyen} is a Senior Lecturer within the School of Information Technology, Deakin University, Victoria, Australia. Nguyen's research interests include Computer Vision, Machine Learning, and Multimedia Computing. He has published his work in highly-ranked publication venues in the field such as Pattern Recognition journal, IEEE Transactions, CVPR, ICCV, ECCV, KDD, and AAAI. He has been an Area Chair of the Multimedia Analysis and Understanding track for the IEEE International Conference on Multimedia and Expo since 2021. He has been a Reviewer for many international journals and a Technical Program Committee Member for many premium conferences in his research field. Nguyen has attracted and managed competitive national/international research funding with a total income over \$2.5 mil AUD.
\end{IEEEbiography}

\begin{IEEEbiography}
[{\includegraphics[width=1in,height=1.25in,clip,keepaspectratio]{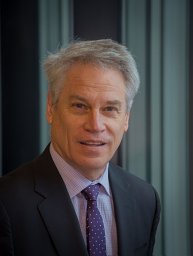}}]{John Yearwood} is a Professor of Computer Science within the School of Information Technology, Deakin University, Victoria, Australia. He was instrumental in setting up the Internet Commerce Security Laboratory with Westpac, IBM and the Victorian State Government as a joint industry-focused and data-driven laboratory on cyber security in the financial sector. He has held a number of ARC grants and was a QEII Fellow working on computational narrative and argumentation in decision science. Professor Yearwood has published over 200 journal and refereed conference papers including 2 books. Professor Yearwood is currently a CI on the ARC funded Discovery Project ``Enhancing and supporting deliberation in multi-disciplinary team decision-making''. He is Editor-in-Chief of the Journal of Research \& Practice in Information Technology and a reviewer for a large number of journals and competitive research grant programs including the Australian Research Council grant program, the NHMRC grant program, and for the Dutch Government in the assessment of their NWO/ToKeN2000.
\end{IEEEbiography}


 





\end{document}